\documentclass[twocolumn,aps,prl,showpacs,superscriptaddress]{revtex4}
\usepackage{graphicx}
\usepackage{amsfonts}
\usepackage{amsmath}
\usepackage{amssymb}

\begin{document}

\title{Bosonic bunching does not reveal stronger quantum contextuality}


\author{Ad\'an Cabello}
 \email{adan@us.es}
 \affiliation{Departamento de F\'{\i}sica
 Aplicada II, Universidad de Sevilla, E-41012 Sevilla, Spain}
 \affiliation{Universidade Federal de Minas Gerais,
 Caixa Postal 702, 30123-970, Belo Horizonte, MG, Brazil}





\date{\today}



\begin{abstract}
In a recent paper, Kurzy\'nski {\em et al.}\ present a gedanken experiment that, they claim, violates the Klyachko-Can-Binicio\u{g}lu-Shumovsky (KCBS) noncontextuality (NC) inequality beyond its maximum quantum value, and a similar experiment that, they claim, violates Specker's NC inequality, which is not violated by quantum mechanics. We argue that these claims are baseless, since these experiments do not satisfy the conditions required for any observation of contextuality through the experimental violation of a NC inequality. Moreover, the physical events in the experiments of Kurzy\'nski {\em et al.}\ do not have the relationships of exclusivity which the authors assume they have.
\end{abstract}


\pacs{03.65.Ud, 42.50.Xa}

\maketitle


{\em Introduction.---}Contextuality is defined as the impossibility of explaining the results of tests (or measurements), assuming that these results do not depend on whether or not other compatible tests are performed. When we talk about compatible tests, it is implicit that each of them is {\em repeatable}: the sequential execution of identical tests on the same physical system must give identical results, no matter how the physical system is initially prepared \cite{Peres95}. Two repeatable tests $A$ and $B$ are {\em compatible} when their sequential execution on the same physical system gives identical results for each test, no matter the order in which they are performed or how the physical system is initially prepared \cite{Peres95}.

Contextuality is detected following a well-established procedure: the experimental violation of a noncontextuality (NC) inequality, which is a condition satisfied by any theory which assumes that results of (repeatable) tests do not depend on whether or not other compatible (repeatable) tests are performed; these theories are called noncontextual hidden variable (NCHV) theories.

A NC inequality usually contains a linear combination of probabilities of physical events, each of them {\em involving only compatible (repeatable) tests}. For example, probabilities such as $P(A=1,B=-1)$, denoting the probability of obtaining the result $1$ when the test $A$ is performed, and the result $-1$ when the compatible test $B$ is performed (after or before $A$; the order is irrelevant since $A$ and $B$ are compatible). A physical event is, for example, $(A=1,B=-1)$, denoting ``the result $1$ is obtained when $A$ is performed, and the result $-1$ is obtained when $B$ is performed''.

{\em Any} experiment revealing contextuality (i.e., detecting the violation of a NC inequality) requires: (i) physical systems {\em prepared in the same state}, (ii) experiments involving {\em only compatible (repeatable) tests}, and (iii) {\em each test to appear in two or more different sets of compatible tests (called ``contexts'')}.

Quantum mechanics (QM) cannot be described in terms of NCHV theories \cite{Specker60,Bell66,KS67}. Moreover, experiments show the violation of NC inequalities in agreement with the predictions of QM \cite{Cabello08,KZGKGCBR09,ARBC09}.

A famous NC inequality is the Klyachko-Can-Binicio\u{g}lu-Shumovsky (KCBS) \cite{KCBS08} which, written as in \cite{BBCGL11} and completed with its maximum quantum violation, reads
\begin{widetext}
\begin{equation}
 \label{KCBS}
 P(A=1,B=-1)+P(B=1,C=-1)+P(C=1,D=-1)+P(D=1,E=-1)+P(E=1,A=-1) \stackrel{\mbox{\tiny{ NCHV}}}{\leq} 2 \stackrel{\mbox{\tiny{ QM}}}{\leq} \sqrt{5},
\end{equation}
\end{widetext}
where $\stackrel{\mbox{\tiny{ NCHV}}}{\leq} 2$ indicates that no NCHV theory can give a value higher than 2, and $\stackrel{\mbox{\tiny{ QM}}}{\leq} \sqrt{5}$ indicates that a value higher than $\sqrt{5} \approx 2.236$ cannot occur in QM. The fact that $\sqrt{5}$ is the maximum value in QM is shown in \cite{KCBS08,BBCGL11}. Indeed, $\sqrt{5}$ is the maximum possible value in any theory satisfying the principle that the sum of the probabilities of pairwise exclusive events cannot exceed 1 \cite{Cabello12}.

In inequality (\ref{KCBS}), there are 5 tests ($A,B,C,D$, and $E$), 5 contexts ($\{A,B\}, \{B,C\}, \{C,D\}, \{D,E\}$, and $\{A,E\}$), and each test appears in two contexts. There are 5 events [$(A=1,B=-1)$, $(B=1,C=-1)$, $(C=1,D=-1)$, $(D=1,E=-1)$, and $(E=1,A=-1)$], and each event only involves two compatible tests. On the other hand, event $(A=1,B=-1)$ and event $(B=1,C=-1)$ are exclusive (i.e., both cannot be true at the same time; see below for further discussion). Similarly, $(B=1,C=-1)$ and $(C=1,D=-1)$ are exclusive, etc. The relationships of exclusivity between the 5 events in (\ref{KCBS}) can be represented by a pentagon, where events are represented by vertices and exclusive events by adjacent (i.e., linked) vertices.

Similarly, for any NC inequality one can construct the graph $G$ of the relationships of exclusivity between the events appearing in that inequality. In \cite{CSW10}, it is shown that the maximum value for NCHV theories is given by the independence number of $G$, and that the maximum value in QM is upper bounded by the Lov\'asz number of $G$. For the pentagon, the independence number is $2$ and the Lov\'asz number is $\sqrt{5}$.


{\em Bosonic bunching and quantum contextuality.---}However, in a recent paper \cite{KSTK12}, Kurzy\'nski {\em et al.}\ claim to have found a value of $\frac{5}{2}$ for a NC inequality such that their relations of exclusivity, they say, are represented by a pentagon. As they note, this result contradicts \cite{Cabello12}. Indeed, it also contradicts \cite{KCBS08,BBCGL11,CSW10}.

Here we argue that the experiment described in \cite{KSTK12} is not a test of the KCBS inequality or any other NC inequality whose relationships of exclusivity are represented by a pentagon (e.g., \cite{SBBC11}).

The first reason is that the tests are not performed on a system prepared in the same state. The second is that none of the events considered involves two or more compatible tests. And the third, that the relationships of exclusivity of the 5 events are not represented by a pentagon, but by 5 isolated vertices.

To illustrate these points, let us focus on two of the 5 events of the experiment in \cite{KSTK12} which, according to Kurzy\'nski {\em et al.}, are exclusive:
\begin{itemize}
\item[] Event 1: Detector ${\cal D}$ clicks when one photon, coming from fiber $A$, is injected in the upper input port of a 50:50 beam splitter.
\item[] Event 2: Detector ${\cal D}$ does not click when two photons, one coming from fiber $A$ and the other from fiber $B$, are simultaneously injected in, respectively, the upper and lower input ports of a 50:50 beam splitter.
\end{itemize}
A first problem is that, in each event, the distinction between which is the state preparation and which are the compatible tests is not clear. In any case, which is the initial state common to both events? Apparently, there are two different states: one with one photon and another with two photons. If the common initial state is ``one photon in fiber $A$'', then injecting a second photon should be part of a test. Then, which is the test?· Which are the other compatible tests? In any of these events, there is neither a way to identify two (potentially repeatable) compatible tests that can be performed sequentially in any order, nor a single repeatable test that is in two different contexts. Consequently, these two events cannot be part of a NC inequality.

Moreover, why are these two events exclusive? Kurzy\'nski {\em et al.}'s explanation is that they are assuming that each of the photons have decided, before entering in the beam splitter, which detector will click. They identify this assumption with the assumption of noncontextuality of results of tests defined above. Why is that if there are no compatible tests involved?

Another problem in \cite{KSTK12} is the definition of exclusive events. Two physical events are exclusive when both cannot be true at the same time; for instance, $(A=1)$ and $(A=-1)$. Two exclusive events, each of them involving only compatible tests, may involve incompatible tests; for instance, if $B$ and $A$ are compatible and $B$ and $C$ are compatible, the two events $(B=-1, A=1)$ and $(B=1,C=-1)$ are exclusive even though $A$ and $C$ may be incompatible. The reason is that their exclusivity can be decided by testing $B$: since the order is irrelevant $(B=-1, A=1)$ may be the event ``the results $-1$ and $1$ are respectively obtained when $B$ and $A$ are sequentially measured (on a physical system prepared in the state $\rho$)'', while $(B=1,C=-1)$ may be the event ``the results $1$ and $-1$ are respectively obtained when $B$ and $C$ are sequentially measured (on a physical system prepared in the state $\rho$)''. To decide exclusivity, it is enough to test $B$. Whether or not two events, each of them involving only compatible tests, are exclusive is {\em experimentally decidable} and {\em does not depend on any particular model of the events}. However, Kurzy\'nski {\em et al.}'s notion of ``exclusivity'' is not testable and only makes sense within a specific model, and not in QM or in more general theories.

The same arguments apply to the analysis of the experiment presented in \cite{KSTK12} which supposedly violates Specker's inequality \cite{Specker60},
\begin{widetext}
\begin{equation}
 \label{S}
 P(A=1,B=-1)+P(B=1,C=-1)+P(C=1,A=-1) \stackrel{\mbox{\tiny{ NCHV, QM}}}{\leq} 1,
\end{equation}
\end{widetext}
whose graph of relationships of exclusivity is a triangle. For the triangle, both the independence number and the Lov\'asz number are $1$.

In this case, Kurzy\'nski {\em et al.}\ assume, for instance, that the following two events are exclusive:
\begin{itemize}
\item[] Event 1': Detector ${\cal D}$ does not click when two photons, one coming from fiber $A$ and the other from fiber $B$, are simultaneously injected in, respectively, the upper and lower input ports of a 50:50 beam splitter.\\
\item[] Event 2': Detector ${\cal D}$ clicks when two photons, one coming from fiber $A$ and the other from fiber $C$, are simultaneously injected in, respectively, the upper and lower input ports of a 50:50 beam splitter.
\end{itemize}
Again, there is neither a way to identify an initial state common to all events, nor tests $a,b$, and $c$ such that $\{a,b\}$ is a context and $\{a,c\}$ is a different context, nor any reason to conclude that these two events are exclusive. Consequently, these two events cannot be part of a NC inequality.


{\em Conclusions.---}In brief: There is a well-established procedure to experimentally reveal contextuality by observing the violation of a NC inequality. None of the experiments in \cite{KSTK12} satisfies the requirements of this procedure. Therefore, nothing in \cite{KSTK12} contradicts any of the results in \cite{KCBS08,BBCGL11,CSW10,Cabello12} or can be interpreted as a proof of stronger quantum contextuality.


\begin{acknowledgments}
 The author thanks C. Budroni, O. G\"uhne, M. Kleinmann, and M. Terra Cunha for useful conversations.
 This work is supported by the Science without Borders Program (Capes and CNPq, Brazil) and the Project No.\ FIS2011-29400 (Spain).
\end{acknowledgments}



\end{document}